\documentclass[aps,prb,twocolumn,groupedaddress]{revtex4}

\usepackage{graphicx}
\usepackage{dcolumn}
\usepackage{bm}

\begin{document}


\title{RANDOM FIELD MODELS FOR RELAXOR FERROELECTRIC BEHAVIOR}


\author{Ronald Fisch}
\affiliation{Department of Materials Science and Engineering\\
University of Pennsylvania\\
Philadelphia, Pennsylvania 19104}


\date{\today}

\begin{abstract}
Heat bath Monte Carlo simulations have been used to study a
four-state clock model with a type of random field on simple cubic
lattices.  The model has the standard nonrandom two-spin exchange
term with coupling energy $J$ and a random field which consists of
adding an energy $D$ to one of the four spin states, chosen
randomly at each site.  This Ashkin-Teller-like model does not
separate; the two random-field Ising model components are coupled.
When $D / J = 3$, the ground states of the model remain fully
aligned.  When $D / J \ge 4$, a different type of ground state is
found, in which the occupation of two of the four spin states is
close to 50\%, and the other two are nearly absent. This means
that one of the Ising components is almost completely ordered,
while the other one has only short-range correlations.  A large
peak in the structure factor $S ( k )$ appears at small $k$ for
temperatures well above the transition to long-range order, and
the appearance of this peak is associated with slow, "glassy"
dynamics.  The phase transition into the state where one Ising
component is long-range ordered appears to be first order, but the
latent heat is very small.

\end{abstract}


\maketitle

\section{INTRODUCTION}

Shortly after the seminal work of Imry and Ma\cite{IM75} on the
effects of random fields in ferromagnets, it was pointed out by
Halperin and Varma\cite{HV76} that similar ideas could be used to
understand the effects of atomic disorder in perovskite
ferroelectrics.  Halperin and Varma showed that atomic disorder
which coupled linearly to the ferroelectric order parameter
behaved in an equivalent fashion to the random field in a
ferromagnet, and thus could cause the ferroelectric phase
transition to become smeared.  They seem to have been unaware,
however, that ferroelectrics displaying such smeared transitions
had been discovered long before.\cite{SI54}  These materials, now
called relaxor ferroelectrics, were originally referred to as
ferroelectrics with diffuse phase transitions.  The random-field
ideas of Halperin and Varma did not immediately become popular in
the field.  In the 1984 review of Isupov\cite{Smo84} the diffuse
phase transitions were modeled using only random-bond disorder.
However, by the early 1990's it was generally recognized that
random-bond disorder was not enough to explain all of the observed
effects, and models including both random bonds and random fields
became widely used.\cite{WKG92,Kle93,PTB87,PB99} Until now it has
been assumed\cite{PTB87,PB99} that in the presence of a strong
cubic crystal-field anisotropy one could treat the different Ising
components independently.  In this work we will show that for a
certain type of random field new phenomena occur which cannot be
understood without including coupling between the components.

The identification of relaxor ferroelectrics as a broad class of
materials in which electric dipoles behave in a manner analogous
in many respects to the magnetic dipoles in spin-glasses was made
forcefully by Burns and Dacol,\cite{BD83} who generalized and
extended ideas of Courtens.\cite{Cou82}  It is essential, however,
to note that, as shown by Halperin and Varma,\cite{HV76} in the
case of electric dipoles it is almost inevitable that the disorder
which is an essential feature of these materials will produce
random fields.  The presence of these random fields limits the
development of long-range spin-glass order in three
dimensions.\cite{HFA79} Therefore, it is not expected that the
glassy behavior which is seen in the disordered electric dipole
materials represents a true phase transition into a state with
long-range spin-glass-type order.

We would like to study the development of spatial correlations
among the electric dipoles.  In order to do this, we will work
with a model which does not retain all of the atomic details.  In
the model studied here, we crudely represent each perovskite unit
cell by only one dynamical variable. We will refer to this
variable as a spin, although it actually represents atomic
displacements in the unit cell.  The model includes a cubic
crystal anisotropy, a nearest-neighbor two-spin interaction, and a
random field.

As discussed by Pirc and Blinc,\cite{PB99} if we assume that the
cubic crystal anisotropy is so strong that each spin points along
one of the [111] directions and the random field is represented by
a simple vector coupling linearly to the spin, then the different
components of the spin act independently.\cite{Suz67}  Thus the
model reduces to a set of decoupled random-field Ising models.  We
may also include randomness in the bonds. (To perform this
decoupling on Eq.~(1), we use a coordinate system rotated by
45$^\circ$.)  This model has been studied in detail over a number
of years. It is believed to be generally well-understood, although
there remains some controversy about the values of the critical
exponents.\cite{Kle02,Ye02}

Similar models have been studied in which the spin components do
not act independently.  Then the behavior does not reduce to that
of an Ising model. For example, there are the three-state Potts
model with a random field,\cite{EB96} and the four-state Potts
model with a random field,\cite{QB96} which are still
qualitatively similar to the random-field Ising model.  When the
random field becomes strong enough, the ground state of the system
breaks up into Imry-Ma domains, and the long-range order is
destroyed.

There is also the cubic model with random anisotropy,\cite{Fis93}
which, when the randomness is strong, shows a new type of domain
phase with long-range order.  This phase may be thought of as one
in which each of the spin components orders independently, but
different parts of the lattice are dominated by different
components.  Thus, in this phase, there is a network of 90$^\circ$
domain walls embedded in the ordered phase.  A somewhat similar
domain state was found by Kartha, Castan, Krumhansl and
Sethna\cite{KCKS91} in a model of shape-memory alloys.

The existence of such a domain state with long-range order has
been proposed\cite{BDKB95,BKB96} as the explanation of relaxor
ferroelectric behavior.  Further, Egami\cite{Eg00} has emphasized
that the network of 90$^\circ$ domain walls is probably an
essential element leading to the large piezoelectric response of
relaxor ferroelectrics.  Therefore, we would like to study a
random-field model which has such a phase.

\section{THE MODEL}

The simplest model which has the desired properties is a
four-state vector Potts model (four-state clock model) with a
random Potts field, on a simple cubic lattice.  The Hamiltonian of
this model is
\begin{equation}
  H = - J \sum_{\langle ij \rangle} {\bf S}_{i} \cdot {\bf S}_{j}
    + h_{r} \sum_{i} \delta_{{\bf S}_{i} , {\bf n}_{i}} \, .
\end{equation}
Each spin ${\bf S}_{i}$ is a dynamical variable which has four
allowed states: (1,0), (0,1), (-1,0) and (0,-1).  Each ${\bf
n}_{i}$ is an independent quenched random variable which assumes
one of these four allowed states with equal probability.  The
$\langle ij \rangle$ indicates a sum over nearest neighbors of a
simple cubic lattice.  The extension to three-component spins is
conceptually straightforward, but will not be discussed in detail
here.

The sign of $h_{r}$ is chosen so that for large positive values
the probability of spin ${\bf S}_{i}$ in state ${\bf n}_{i}$ is
strongly suppressed.  Density-functional calculations for a
typical relaxor ferroelectric alloy\cite{CGMR02} have shown that
the most significant interactions created by the alloy disorder
are repulsive and short-ranged. Thus we believe that this is a
reasonable first approximation to the interactions produced by the
disorder.

When $h_{r}$ is negative, the random Potts field favors one state
on each site, just as a vector field would.  In that case there is
no qualitative difference between the random Potts field and the
random vector field.\cite{GH96,Fis97}  Thus we will not present
any calculations for negative $h_{r}$ here.

When $h_{r}$ is large and positive, we can approximate the random
field term as a projection operator which forbids ${\bf S}_{i}$
from occupying the state ${\bf n}_{i}$.  Then within mean-field
theory it becomes simple to calculate the approximate ground
states.  There are eight of these mean-field ground states.  For
example, in one such ground state 75\% of the spins are in state
(1,0), and the other 25\% of the spins, which are prevented from
being in this spin state by their local ${\bf n}_{i}$ are in state
(0,-1).  The energy of this state for a simple cubic lattice is
easily calculated to be $-1.875~J$ per spin.  Additional details
of the mean-field theory are discussed in the Appendix.

As we will see, the energies of the ground states found by
computer calculation are less than $-2~J$, and they differ
qualitatively from the mean-field ground states.  The problem with
the mean-field ground states is easy to understand.  If we
consider the above example, on a simple cubic lattice the 25\% of
spins in the state (0,-1) are broken up into finite clusters.  If
we move all of the spins in such a finite cluster into the state
(0,1), the total energy does not change.  One might expect,
therefore, that for large $h_{r}$, the long-range order in the
ground state would still be along one of the four [1,0]
directions, since the 25\% of the spins in the finite clusters
should not be able to exhibit long-range order.

However, a true ground state on the simple cubic lattice actually
has a very different structure.  It turns out to be possible to
find states with, for example, 50\% of the spins in state (1,0)
and 50\% of the spins in state (0,-1) which have a much lower
energy than the mean-field ground states.  As we will show, to
find states which look like the true ground states for large
$h_{r}$ within mean-field theory, we need to add another term to
Eq.~(1).

The author finds it helpful to compare the model studied here to
the results for the three dimensional Ashkin-Teller
model,\cite{DBGK80} whose Hamiltonian is
\begin{equation}
  H_{AT} = - J \sum_{\langle ij \rangle} ({\bm \sigma}_{i} \cdot
   {\bm \sigma}_{j} + {\bm \tau}_{i} \cdot {\bm \tau}_{j})
    - 2 J_{4} \sum_{\langle ij \rangle} ({\bm \sigma}_{i} \cdot
     {\bm \sigma}_{j})({\bm \tau}_{i} \cdot {\bm \tau}_{j}) \, ,
\end{equation}
where ${\bm \sigma} = \pm {1 \over 2} (1,1)$ and ${\bm \tau} = \pm
{1 \over 2} (1,-1)$ are Ising variables.  (Note that our notation
and normalization differ from those of Ditzian {\it et
al.}\cite{DBGK80})

Each of the three linearly independent components of a random
field of a four-state model can now be identified with one of the
terms of Eq.~(2), and thus one of the mean-field order parameters.
Expressing the random-field term using the ${\bm \sigma}$ and
${\bm \tau}$ variables, and adding it to Eq.~(2), we get
\begin{equation}
  H_{RFAT} = H_{AT} + ({h_{r} / 4}) \sum_{i} [ 1
+ 2 ({\bm \sigma}_{i} \cdot {\bf n}_{i}) + 2 ({\bm \tau}_{i} \cdot
{\bf n}_{i}) + 4 \eta ({\bm \sigma}_{i} \cdot {\bf n}_{i})({\bm
\tau}_{i} \cdot {\bf n}_{i})]  \, .
\end{equation}
If we set $\eta = 1$ and $J_{4} = 0$, then Eq.~(3) becomes just
Eq.~(1) expressed in the coordinate system defined by ${\bm
\sigma}$ and ${\bm \tau}$.  If we then set $\eta = 0$, the random
Potts field term is replaced by the usual random vector field
term.

In a real alloy we would not expect that exactly one of the local
states was blocked on each site.  Instead, we would expect to find
a random distribution, with some sites having no blocked states,
some with one blocked state, etc.  It would be straightforward to
generalize the random field term to allow this, by letting $\eta$
in Eq.~(3) become a function of {\it i}.  For example, we could
use a probability distribution for $\eta_{i}$ which had weight at
$\pm 1$.  Such a generalization is likely to be useful in modeling
the properties of specific alloy systems. However, a more general
model has more parameters. Exploring its entire parameter space
would be tedious, and seems premature at this time.

\section{PHASE DIAGRAM}

If $h_{r} / J$ is chosen to be an integer, then the energy of any
state is an integer multiple of $J$.  Then it becomes
straightforward to design a heat bath Monte Carlo computer program
to study Eq.~(1) which uses integer arithmetic for the energies,
and a look-up table to calculate Boltzmann factors.  This
substantially improves the performance of the computer program,
and was used for all the calculations reported here.  (If desired,
one could do almost as well for half-integer values of $h_{r} /
J$.) Lattices with periodic boundary conditions were used
throughout.  Two different linear congruential random number
generators were used: one to choose the random ${\bf n}_{i}$ and
one to choose the Boltzmann factors for the Monte Carlo spin
flips.

A series of $L \times L \times L$ samples of various lattice sizes
and values of $h_{r}$ was studied to map out the phase diagram,
which is shown in Fig.~1.  For $h_{r} / J \le 3$ the behavior of
the random Potts field model remains generally similar to the
behavior of the random vector field model.  At low temperature the
system develops long-range order, with the order parameter aligned
along one of the spin-coordinate axes.  It is not required,
however, that the critical exponents for the phase transition here
will be those of the random-field Ising model, as happens for the
vector random-field case.

The $\langle {\bm \sigma} \rangle$ phase found by Ditzian,
Banavar, Grest and Kadanoff\cite{DBGK80} has many of the features
of the domain FE phase which we find in our model for large
$h_{r}$ and low $T$ . The $\langle {\bm \sigma} \rangle$ phase of
Eq.~2 only occurs for $J_{4} < 0$. This indicates that one of the
effects of the random Potts field in Eq.~(1) is to generate a
negative effective value of the $J_{4}$ coupling.  This effect was
also seen in the random-anisotropy cubic model.\cite{Fis93}  It is
the random-anisotropy part (the $\eta$ term of Eq.~(3)) which
generates the $J_{4}$ effective coupling under rescaling, and
prevents the decoupling of Eq.~(1) into two Ising models. If we
want to study a mean-field theory for Eq.~(1) we should include a
$J_{4}$ term, thus arriving at Eq.~(3).  The effective value of
$J_{4}$ will depend on $h_{r}$. In this way we can obtain a
mean-field theory which reproduces the phase diagram of Fig.~1.

A necessary condition for the existence of stable long-range [1,1]
domain FE order is that there be infinite connected clusters of
both of the two dominant spin states.  Since it becomes impossible
to fulfill this condition when the amount of long-range order is
small, the transition from the domain FE phase to the paraelectric
phase must be first order.  This, however, is specific to our
model with only first neighbor interactions on the simple cubic
lattice.  It is possible that this phase transition could become
continuous if, for example, second neighbor exchange is included
in the Hamiltonian.

As discussed by Ditzian {\it et al.}\cite{DBGK80}, even in the
absence of randomness the details of what happens as we move from
the $\langle {\bm \sigma} \rangle$ phase to the Baxter phase are
not clear. There may be a small area in the region of the [1,0] FE
to [1,1] domain FE phase boundary in which an intermediate phase
is stable.  This would be similar to the intermediate phase which
has recently been found experimentally.\cite{NCSGCP99}  We have
indicated the uncertainty about what is going on in this region of
the phase diagram by the question mark in Fig.~1. A generalization
of the $\langle {\bm \sigma} \rangle$ phase also exists for the
three-component Ashkin-Teller model in three
dimensions.\cite{GW81}

The dotted line in Fig.~1 represents the approximate location of
the onset of thermal hysteresis.  Above the dotted line, the
system will quickly relax to a state which is independent of
initial conditions, but below this line the system retains a
memory of initial conditions for a long time. This line does not
represent any true thermodynamic singularity.  Its precise
location depends somewhat on the size of $L$ and the length of the
Monte Carlo runs.  Thus the dotted line should be identified as
the glass temperature, $T_{g}$.  For small values of $L$ the
system can develop order on the scale of $L$ above $T_{g}$.  For
$L = 64$ and $h_{r} \ge 3$ it is not possible to find the ground
states in a reasonable time by cooling the system from high
temperature.

\section{SPECIFIC HEAT}

In Fig.~2 we display the specific heat for Eq.~(1) using four
different values of $h_{r}$.  These curves were calculated by
numerically differentiating and averaging Monte Carlo data for the
energy of runs for four $L = 64$ lattices for each value of
$h_{r}$, starting at high temperature and cooling slowly.  For
each run, the temperature was changed in steps of $0.05 J$. 40,960
Monte Carlo passes through the lattice were performed at each
value of $T$, with data being collected after each 20 passes. The
first 248 data points at each temperature were discarded, and the
remaining 1,800 data points were averaged.  A similar procedure
was used for heating runs.

Note that for $h_{r} = 0$ (not displayed), the specific heat for
Eq.~(1) reduces to a that of an ordinary Ising model on the simple
cubic lattice, with $T_c / J = 2.256$.  This model has a very
sharp singularity at $T_c$, whose height diverges to infinity with
$L$.  We see in Fig.~2 that as $h_{r}$ increases, the peak in the
specific heat broadens and shifts to lower energy, with
approximately half of the change occurring by $h_{r} = 3$.  By
comparing with Fig.~1, it is seen that for $h_{r} > 0$ the peak in
the specific heat is centered above the region of glassy behavior.
The average heights and widths of the peaks shown in Fig.~2 are
essentially the same for $L = 32$ (not shown) as for $L = 64$.  As
one would expect, there is more variation from sample to sample
for the smaller lattices.

In Fig.~3 we compare data from the cooling runs with $h_{r} = 6$
with heating runs on the same lattices.  The initial condition for
each heating run was a state with long-range order.  The
temperature at which the long-range order collapses is identified
as $T_{g}$.  We see that below $T_g$ there is a region of $T$
where the specific heat is slightly higher for the cold-start
sample. The crossing of the two curves near $T/J = 1.2$ gives us a
crude estimate for the ordering temperature $T_c$.

It is often possible to estimate $T_c$ for a first-order phase
transition with high precision.  In order to do this with a
computer simulation, however, it is necessary to run the
simulation close to $T_c$, with $L$ larger than the correlation
length, for a time which is long enough so that the sample can
spontaneously order and disorder several times.  We are far from
being able to satisfy that condition here, due to the glassy
dynamics for $T < T_g$.

The integrated area between the cooling curve data and the heating
curve data gives us an estimate of the value of the latent heat at
$T_c$. We can also look directly at the difference in energy
between the heating run and the cooling run at $T = 1.2~J$.  Thus
we find that the latent is about $0.01~J$.  The fact that the
slope of the heating curve is so similar to that of the cooling
curve and the latent heat at $T_c$ is so small indicates that
there is very little difference in the local structure of this
model between the ordered states of the [1,1] domain ferroelectric
phase and the metastable states which are found by slow cooling. A
small latent heat is a natural consequence of a large (but finite)
correlation length at $T_c$.

\section{CORRELATION FUNCTIONS}

In order to obtain information about the two-spin correlations, we
calculate the angle-averaged structure factor $S ( k )$.  In
Fig.~4 we display a log-log plot of $S ( k )$, obtained by
averaging over the data from the cooling runs for the $L = 64$
lattices, as a function of $T / J$.  We see that for all these
values of $h_{r} \ge 3$, the qualitative behavior of a sample
cooled from high $T$ without any external ordering field is the
same. As we lower $T$, the peak near $k = 0$ first grows and then
saturates.  The rate of growth of this peak is maximal at the
temperature where the specific heat has its maximum.  As $h_{r}$
decreases, the height of the peak and the value of the correlation
length increase.  For $h_{r} = 3$, the correlation length at low
$T$ becomes larger than our sample size.  However, these
zero-field-cooled samples never show true long-range order.  With
true long-range order, the size of the small-$k$ peak would
decrease as $T$ decreases below $T_c$, because the intensity in
the long-range-order $\delta$-function will not appear on the
log-log plot.

In Fig.~5 we show data for $S ( k )$ from the heating runs, where
the samples begin in an ordered state.  Under these conditions,
the size of the small-$k$ peak continues to grow as $T$ is
increased above $T_c$, until $T$ reaches $T_{g}$ and the sample
can equilibrate.  In Fig.~5(c) we show data from runs which are
ten times longer than the standard runs.  The $h_{r} = 4$ samples
are able to equilibrate at $T / J = 1.7$ during the long run ({\it
i.e.} they become indistinguishable from the data from the
corresponding cooling run), but the changes at $T / J = 1.5$ are
small.  By $T / J = 1.4$ increasing the length of the run by a
factor of ten has a negligible effect on the observed state of the
sample.

\section{GROUND STATES}

Somewhat surprisingly, it turned out to be possible to find
approximate ground states of the $h_{r} = \infty$ samples by a
simulated annealing procedure.  In Fig.~6 we show $S ( k )$ for
the approximate ground states of the $L = 64$ lattices with $h_{r}
= \infty$, averaging over each of the four [1,1] directions for each
of the four samples.  The data shown in the figure are fit quite
well with by a Lorentzian line shape, with a correlation length of
about 10 lattice units.  This diffuse-scattering part of $S ( k )$
contains about 54\% of the spectral weight.  The other 46\% of the
weight is in the $\delta$-function at $k = 0$, which is due to the
long-range [1,1] order.

What this means in detail is that for large $h_{r}$ a ground state
contains approximately 49\% each of the two majority spin states,
and about 1\% each of the two minority states.  The energy per
spin of the lowest ground states found for each sample was about
$-2.010~J$, with the energies of the "ground states" of the same
sample in the other three [1,1] directions lying typically about
$0.003~J$ higher. Since $L = 64$ is large compared to a
correlation length of 10, this result for the ground state energy
should represent the large $L$ limit, while the differences in
energies between alternate ground state directions of the same
sample should scale like $1 / L^{3/2}$.  Note that these actual
ground state energies are substantially below the energy of the
simple mean-field ground state of the $h_{r} = \infty$ model.

For $L = 64$ and $h_{r} = \infty$ the energy of a
zero-field-cooled state at $T = 0$, which has no long-range order,
is only about $0.01~J$ higher than the energy of a true ground
state.  Thus for lattices smaller than about $L = 32$ the energy
differences between the ground states of different [1,1]
directions become larger than the energy difference between a
zero-field-cooled state and a true ground state.  Under these
conditions the zero-field-cooled sample is able to reach a
long-range-ordered state. The zero-field-cooled $L = 64$ samples
can be decomposed into domains of different [1,1] ground states,
with 90$^\circ$ walls between the domains.

The probabilities of having effective fields of magnitude 6, 5,
... and 0 in a ground state, averaged over the four ground states
for each of these $L = 64$ $h_{r} = \infty$ samples are 0.1121,
0.2780, 0.3148, 0.1685, 0.0751, 0.0383, and 0.0131, respectively.
Thus, in a ground state only about one spin in nine is surrounded
by fully aligned nearest neighbors.  More than 1\% of the spins
are in zero effective field.

Some of the spins can flip freely between two positions in a
ground state with no cost in energy.  Since the spins in this
model are discrete and two-dimensional, being in zero effective
field is not a necessary condition for a spin to be "free" in this
way.  The bulk of these free spins are flipping between the two
majority spin states for that particular ground state.  It is easy
to find that the fraction of free spins in a particular ground
state is about 6.5\%.  However, a spin which is not free in one
ground state may become free by the flipping of other free spins.
The fraction of spins which can become free by the flipping of
other free spins, one at a time, is 11.9\%.  Since 0.119 is much
less than the critical concentration for uncorrelated site
percolation on a simple cubic lattice, it is not surprising that a
set of free spins defined in this way consists of small isolated
clusters.

The probability that a free spin may occupy all three of its
allowed spin states with no energy cost is negligible to this
level of accuracy.  Thus a lower bound on the ground state entropy
per spin for large $h_{r}$ is $0.065~ln (2)$, and $0.119~ln (2)$
is an upper bound.  Calculating the ground state entropy of this
model is a complex problem, because it requires finding all of the
ground states.

Thus each of the four "[1,1] ground states" of a sample is
actually a cluster of ground states in the phase space of the
model.  The simulated annealing procedure works because, if we use
one of the simple mean-field ground states, which has 75\% of the
spins in one spin state and 25\% of the spins in a second spin
state, as an initial condition, the direction of the order
parameter will rapidly relax to the closest of the [1,1] energy
minima, even at temperatures well below $T_c$.  The fact that this
is possible is partly due to the large number of spin
rearrangements which can be made with no energy cost.  This
prevents the system from being easily trapped in a metastable
minimum of the free energy which is close to the initial
mean-field ground state.

Although a ground state energy of $-2.01~J$ means that about 33\%
of nearest neighbor spin pairs are oriented at 90$^\circ$ from one
another, the fraction of nearest neighbor spin pairs which are
pointing in opposite directions in one of these ground states is
only 0.00011. The small number of spins in the minority states
exist in compact blobs whose diameter is approximately the
correlation length of 10 lattice units.

For $h_{r} / J$ = 3, every sample studied had four ground states,
each one fully aligned along one of the four [1,0] spin states.
There is an exponentially small probability of having an unusual
local configuration of the ${\bf n}_{i}$ which would misalign a
small cluster of spins in a ground state, but this was never
observed. Thus, ignoring the statistical fluctuations in the
sample average of ${\bf n}_{i}$, the ground state energy in the
[1,0] ferroelectric phase is found to be
$E_0~=~-3~J~+~0.25~h_{r}$. The value of $E_0$ for any value of
$h_{r}$ cannot be greater than $-2.01~J$, its value for $h_{r} =
\infty$. From this we expect that the boundary between the [1,0]
phase and the [1,1] domain phase should be at $h_{r} / J$ slightly
less than 4, as observed.

\section{DISCUSSION}

In the past, various kinds of evidence have been presented to
argue that relaxor ferroelectrics represent a distinct class of
materials, and are not merely the inevitable consequence of adding
some alloy disorder to any ferroelectric.  De Yoreo, Pohl and
Burns\cite{DPB85} studied the low-temperature properties of a
variety of ferroelectric alloys, and found that they could be
separated into two classes.  Relaxor ferroelectrics showed glassy
low temperature behavior and normal ferroelectrics did not, even
when the normal ferroelectrics were random alloys. Recently,
Viehland {\it et al.}\cite{DXV94,XKLV96,VPCL01} have argued that
in some materials they find a well-defined transition between a
normal ferroelectric phase and a relaxor ferroelectric phase. It
would be remarkable if the low temperature thermal properties of
such a crystal could be switched from glassy behavior to
crystalline behavior by poling the sample.  To the author's
knowledge, this experiment has not yet been tried.

For the two-spin exchange interaction of Eq.~(1), a 180$^\circ$
domain wall has twice the energy of a 90$^\circ$ domain wall.  As
a result of this, 180$^\circ$ domain walls become rare at low
temperature in this model.  As pointed out by B\"urgel, Kleemann
and Bianchi,\cite{BKB96} ferroelastic interactions raise the
energy of 90$^\circ$ walls, but do not have much effect on the
energy of 180$^\circ$ walls.

To incorporate this effect into our model, we would add a
biquadratic term to Eq.~(1) of the form $- K ({\bf S}_{i} \cdot
{\bf S}_{j})^2$. When $K / J$ becomes large, the structure of the
domain state will change,\cite{BKB96} because 180$^\circ$ domain
walls now have a lower energy than 90$^\circ$ domain walls.  It is
difficult to study this B\"urgel, Kleemann, Bianchi domain state
with computer simulations if the size of a ferroelastic domain is
large. A sample which is the size of a single ferroelastic domain
(or smaller) will behave essentially as a random-field Ising
model.

\section{CONCLUSION}

In this work we have used Monte Carlo computer simulations to
study a simple model of a ferroelectric alloy.  In this model, the
alloy disorder causes some of the positions of the polarizable
ions to be strongly disfavored.  We have found that this model
displays a type of ferroelectric domain phase which does not exist
in the model of Pirc and Blinc.\cite{PB99}  This phase may be
thought of as resulting from the addition of a random field to the
$\langle {\bm \sigma} \rangle$ phase\cite{DBGK80} of an Ashkin-Teller
model.  It seems likely that this model, and some natural
generalizations of it, will help in the understanding of relaxor
ferroelectric behavior.

\appendix*
\section{MEAN FIELD THEORY}
The presence of the random field greatly complicates the
mean-field theory for Eq.~(1).  It is necessary to distinguish the
four classes of spins corresponding to the four different allowed
values of ${\bf n}$.  Label the four directions (1,0), (0,1),
(-1,0) and (0,-1) as direction 1, 2, 3 and 4, respectively. Define
$p_{\mu \nu}$ to be the probability that a spin of class $\mu$
points in the $\nu$ direction.  Then $\sum_\nu p_{\mu \nu} = 1$
for any $\mu$.  But, in the general case, we have to deal with
twelve independent variables in the mean-field theory.  If we take
the limit that $h_r$ becomes infinite, there are only eight
independent variables, because we can then set $p_{\mu \nu}$ to 0
when $\mu = \nu$.

Let $z$ be the number of neighbors of any spin.  Then the
mean-field expression for the energy per spin is
\begin{equation}
  E = - (z J / 32) \sum_{\mu, \nu, \mu^\prime, \nu^\prime} p_{\mu \nu}
  p_{\mu^\prime \nu^\prime} \cos ( \pi (\nu - \nu^\prime) / 2) \, .
\end{equation}
In addition to the usual factor of 1/2 to avoid double-counting,
there is factor of 1/16 arising from the assumption that each
value of $\mu$ ({\it i.e.} each ${\bf n}$) is equally weighted.
The entropy per spin is
\begin{equation}
  S = - 1/4 \sum_{\mu, \nu} p_{\mu \nu} \ln ( p_{\mu \nu} ) \, ,
\end{equation}
and the free energy is, as usual, $F = E - TS$.  Then solving the
mean-field theory requires finding the minimum of $F ( T )$ in the
phase space of the $p_{\mu \nu}$.

\begin{acknowledgments}
The author is grateful to T. Egami, I-W. Chen, S. Vakhrushev, A.
B. Harris, I. Grinberg, and A. M. Rappe for helpful discussions.
This work was supported by the Center for Piezoelectrics by Design
under ONR Grant N00014-01-1-0365.

\end{acknowledgments}


\newpage
\begin{figure}
\caption{\label{fig1} Phase diagram of the random Potts field
model on simple cubic lattices, showing the paraelectric (PE),
ferroelectric (FE), and domain FE phases. The solid lines indicate
first-order transitions, and the dashed line indicate the
approximate onset of glassy dynamics in the PE phase.  The
question mark is discussed in the text.}
\end{figure}

\begin{figure}
\caption{\label{fig2} Specific heat vs. temperature for the random
Potts field model on $64 \times 64 \times 64$ simple cubic
lattices, for various values of $h_{r}$.  Data from
zero-field-cooled runs, averaging data from four samples.}
\end{figure}

\begin{figure}
\caption{\label{fig3} Specific heat vs. temperature for the random
Potts field model on $64 \times 64 \times 64$ simple cubic
lattices, for $h_{r} = 6$, averaging data from four samples.  The
solid line shows zero-field-cooled data using a random initial
condition, and the dashed line shows zero-field-heated data using
an ordered initial condition.}
\end{figure}

\begin{figure}
\caption{\label{fig4} Angle-averaged magnetic structure factor at
a sequence of temperatures for the random Potts field model on $64
\times 64 \times 64$ simple cubic lattices, log-log plot.  The
points show averaged data from four samples, using
zero-field-cooling and a random initial condition. (a) $h_{r} =
\infty$; (b) $h_{r} = 6$; (c) $h_{r} = 4$; (d) $h_{r} = 3$.}
\end{figure}

\begin{figure}
\caption{\label{fig5} Angle-averaged magnetic structure factor at
a sequence of temperatures for the random Potts field model on $64
\times 64 \times 64$ simple cubic lattices, log-log plot.  The
points show averaged data from four samples, using
zero-field-heating and an ordered initial condition. (a) $h_{r} =
6$; (b) $h_{r} = 4$; (c) $h_{r} = 4$, relaxed (see text); (d)
$h_{r} = 3$.}
\end{figure}

\begin{figure}
\caption{\label{fig6} Angle-averaged magnetic structure factor at
$T = 0$ for the random Potts field model with $h_{r} = \infty$ on
$64 \times 64 \times 64$ simple cubic lattices, log-log plot.  The
points show averaged data from four samples, using four states
from each sample, one from each [1,1] direction.}
\end{figure}

\newpage
\begin{figure}
\includegraphics{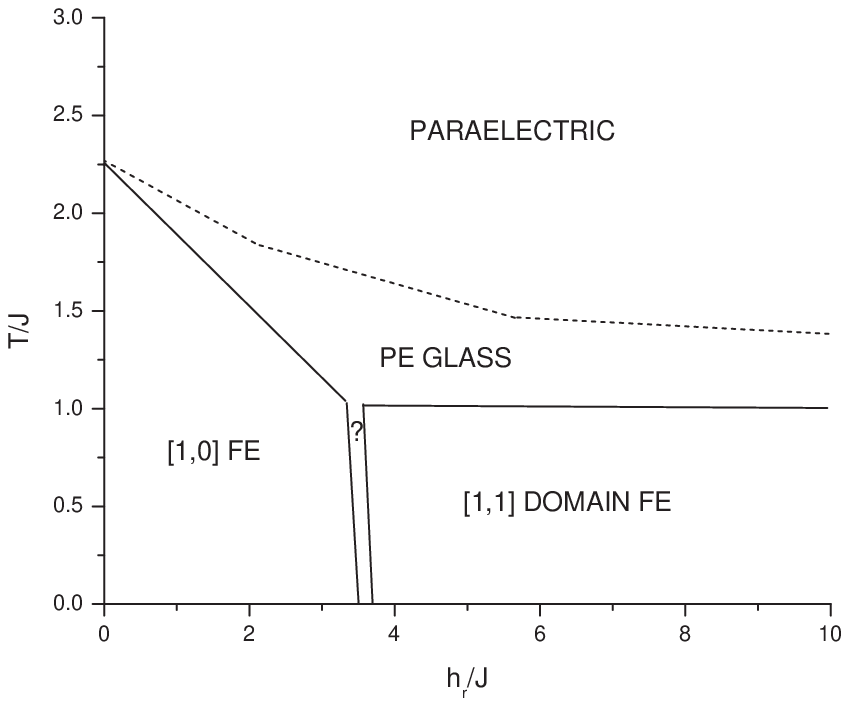}%
\end{figure}

\newpage
\begin{figure}
\includegraphics{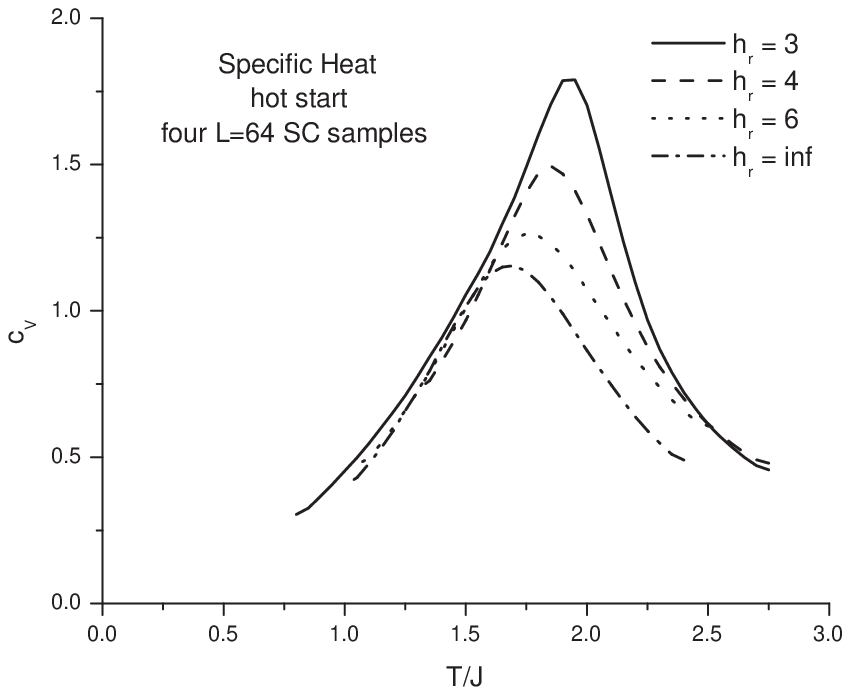}%
\end{figure}

\newpage
\begin{figure}
\includegraphics{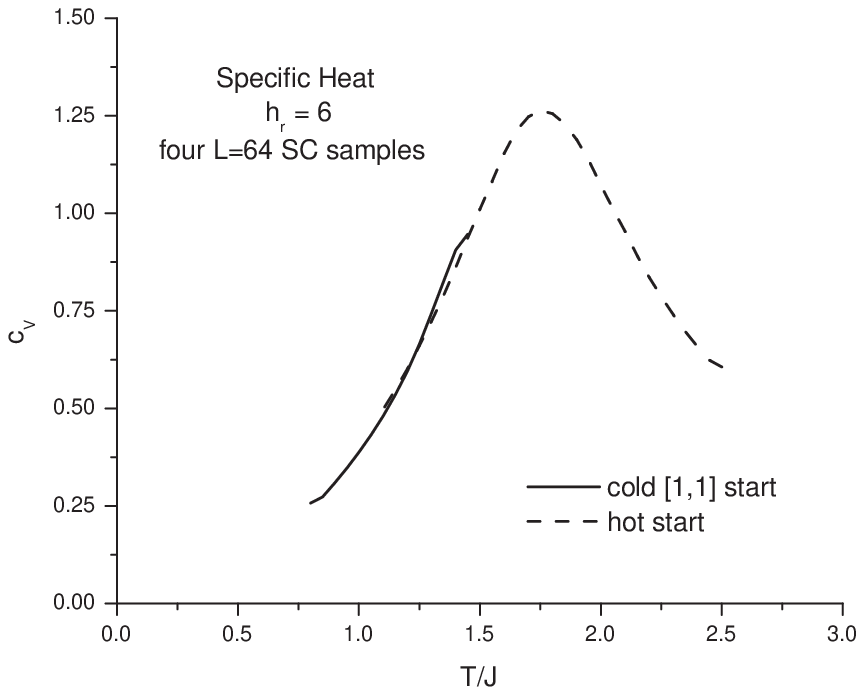}%
\end{figure}

\newpage
\begin{figure}
\includegraphics{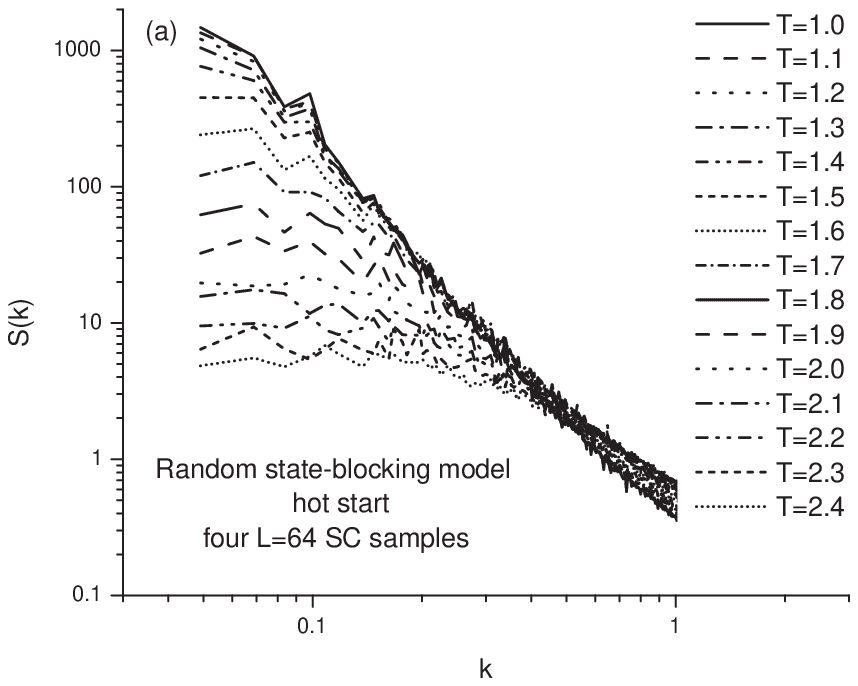}%
\newline
\includegraphics{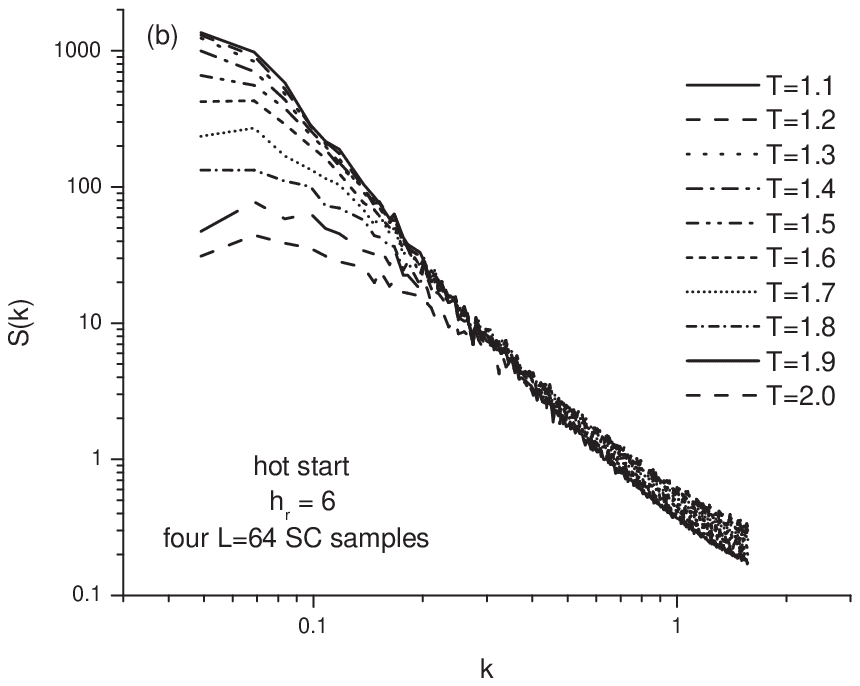}%
\end{figure}

\begin{figure}
\includegraphics{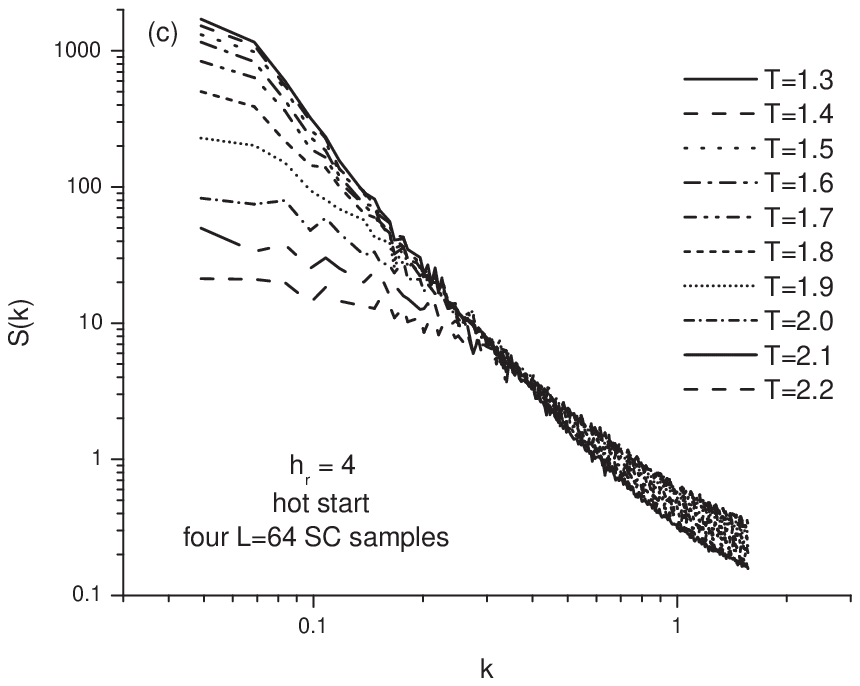}%
\newline
\includegraphics{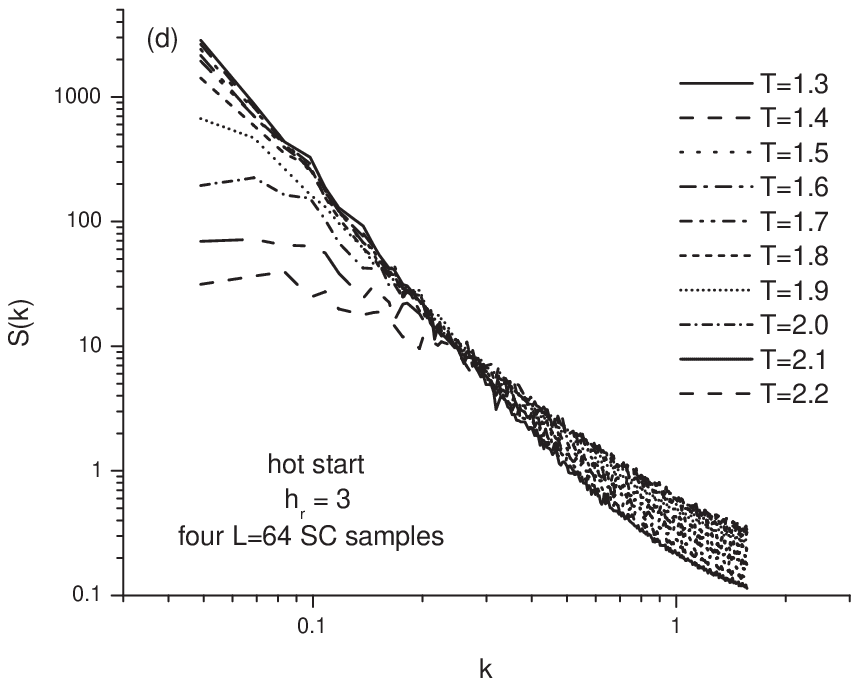}%
\end{figure}

\newpage
\begin{figure}
\includegraphics{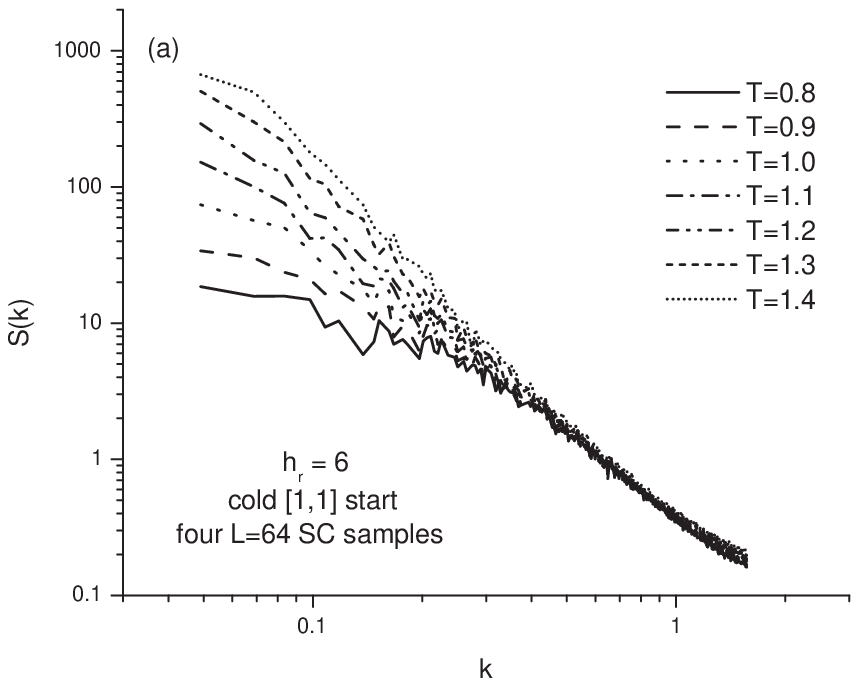}%
\newline
\includegraphics{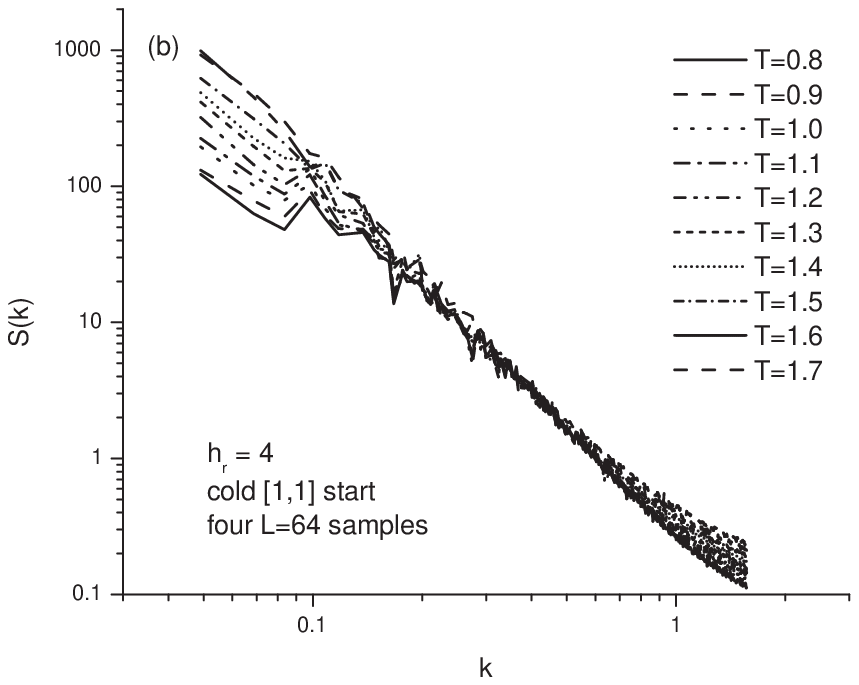}%
\end{figure}

\begin{figure}
\includegraphics{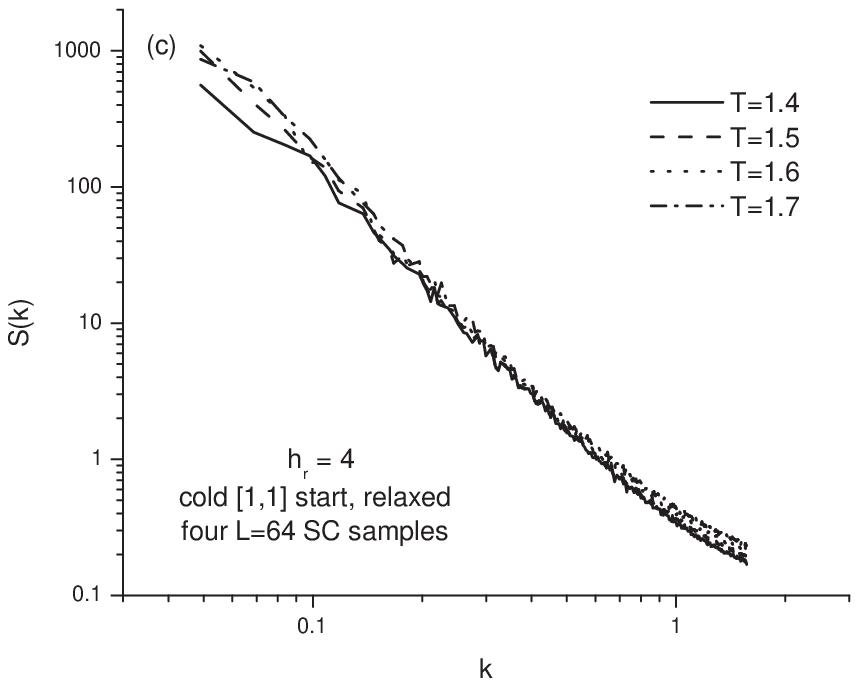}%
\newline
\includegraphics{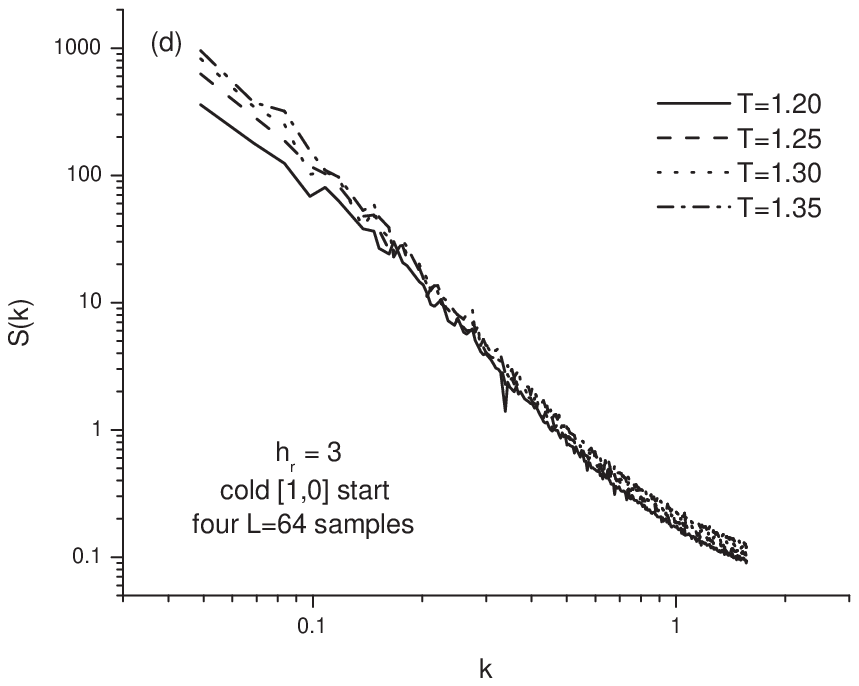}%
\end{figure}

\newpage
\begin{figure}
\includegraphics{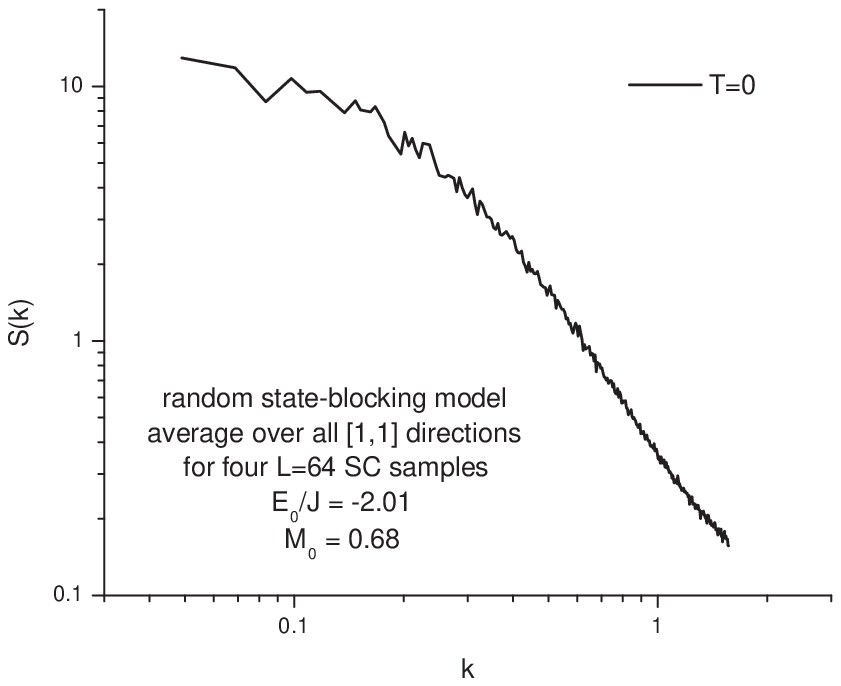}%
\end{figure}

\end{document}